\documentclass[a4paper]{iopart}

\usepackage{latexsym}
\usepackage{amssymb}
\usepackage{color}
\usepackage{ulem}
\usepackage{times}
\usepackage{colordvi}

\usepackage{ifpdf}
\usepackage{graphicx}
\newcommand{\imgname}[1]{#1.eps}

\newcommand{\apj}{Astrophys. J.}
\newcommand{\dr}{{\mathrm d}}

\ifpdf
\fi
\ifpdf
  \renewcommand{\imgname}[1]{#1.pdf}
  
\fi
\begin{document}

\title[Dynamical non-axisymmetric instabilities in rotating
  relativistic stars]{Dynamical non-axisymmetric instabilities in
  rotating relativistic stars}

\author[G.~M. Manca et al.]{Gian Mario Manca,$^{1,2}$ Luca
    Baiotti,$^{3}$ Roberto De Pietri,$^{1}$ and Luciano Rezzolla$^{3,4}$}

\address{$^1$ Dipartimento di Fisica, Universit\`a di Parma and INFN,
  Parma, Italy}

\address{$^2$ Max-Planck-Institut f\"ur Astrophysik, Garching, Germany}

\address{$^3$ Max-Planck-Institut f\"ur Gravitationsphysik,
  Albert-Einstein-Institut, Golm, Germany}

\address{$^4$ Department of Physics and Astronomy, Louisiana State
  University, Baton Rouge, LA, USA}

\date{\today}
\begin{abstract}
  We present new results on dynamical instabilities in rapidly
  rotating neutron-stars. In particular, using numerical simulations
  in full General Relativity, we analyse the effects that the stellar
  compactness has on the threshold for the onset of the dynamical
  bar-mode instability, as well as on the appearance of other
  dynamical instabilities. By using an extrapolation technique
  developed and tested in our previous study~\cite{Baiotti:2006wn}, we
  explicitly determine the threshold for a wide range of compactnesses
  using four sequences of models of constant baryonic mass comprising
  a total of 59 stellar models. Our calculation of the threshold is in
  good agreement with the Newtonian prediction and improves the
  previous post-Newtonian estimates. In addition, we find that for
  stars with sufficiently large mass and compactness, the
  $m\!\!=\!\!3$ deformation is the fastest growing one. For all of the
  models considered, the non-axisymmetric instability is suppressed on
  a dynamical timescale with an $m\!\!=\!\!1$ deformation dominating
  the final stages of the instability. These results, together with
  those presented in~\cite{Baiotti:2006wn}, suggest that an
  $m\!\!=\!\!1$ deformation represents a general and late-time feature
  of non-axisymmetric dynamical instabilities both in full General
  Relativity and in Newtonian gravity.
\end{abstract}

%\maketitle

%-----------------------------------------------------------------
\section{Introduction}
\label{sec:intro}
 %-----------------------------------------------------------------

Non-axisymmetric deformations of rapidly rotating self-gravitating
bodies are rather generic phenomena in nature and could appear in a
variety of astrophysical scenarios like stellar core
collapse~\cite{ShibataSekiguchi2004}, accretion-induced collapse of
white dwarfs \cite{2006ApJ...644.1063D}, or the merger of two neutron
stars~\cite{Shibata:2005ss}. Over the years a considerable amount of
work has been devoted to the search of unstable deformations that,
starting from a quasi-axisymmetric stellar configuration, would lead
to the formation of highly deformed rotating massive objects (see
ref.~\cite{Baiotti:2006wn} for a detailed list of references). One of
the main reasons behind this interest is that such deformations would
lead to the intense emission of high-frequency gravitational waves
(\textit{i.e.}, in the kHz range) which is potentially detectable by
ground-based detectors such as LIGO, GEO, Virgo or the planned
resonant detector such as DUAL~\cite{Bonaldi:Dual2003}.

Despite such extensive studies, various questions about the dynamics
of the non-axisymmetric deformation of rapidly rotating
self-gravitating bodies are not yet completely clarified. Among the
most important questions that have been addressed only rather recently
it is worth to recall the following ones: {\it i)} How long do these
deformation survive once they reach their maximum amplitude? {\it ii)}
How large is the energy emitted in gravitational waves? {\it iii)}
Which physical phenomena determine the shortest damping timescale and
impress a signature on the emitted signal? {\it iv)} What is the
effect of the stellar compactness $M/R_e$, where $M$ and $R_e$ are the
stellar mass and the proper equatorial radius, respectively, on the
dynamics of the instability and on the threshold for its onset?

While points {\it i)}--{\it iii)} were first addressed in
ref.~\cite{Baiotti:2006wn} (hereafter paper I), here we concentrate on
providing an answer to question {\it iv)} supplying new information on
the general properties of the dynamical instability in a very large class
of stellar models that are characterized by differential rotation and by
high compactness and that are members of four sequences of models with
constant baryonic mass.

The main result obtained is that any non-axisymmetric deformation that
develops in our models is damped over a dynamical timescale, through
pure inviscid hydrodynamical nonlinear phenomena. Moreover, for all
the models that develop dynamical instabilities, the $m\!\!=\!\!1$
deformation eventually becomes the dominant one irrespective of
whether the models are above or below the threshold for the
development of the bar-mode instability (see also paper I). This
evidence is consistent with the simulations performed in Newtonian
gravity by Ou and Tohline~\cite{2006ApJ...651.1068O} for stars with a
very-low $\beta$ (where $\beta \equiv T/|W|$ is the ratio between the
rotational kinetic energy $T$ and the gravitational binding energy
$W$), thus suggesting that this may be a general feature of this type
of dynamical instabilities. In addition we show that, in a region of
high stellar compactness, other instabilities, such as one having an
$m\!\!=\!\!3$ deformation, can develop.  Finally, adopting an
extrapolation technique developed and tested in paper I, we determine
the threshold $\beta_c$ for the onset of the bar-mode instability for
all of the sequences considered, thus determining accurately its
dependence on the stellar compactness.

The paper is organised as follows. In Sect.~\ref{sec:initial} we
briefly describe the initial data chosen and the numerical techniques
employed for their evolution, while in Sect.~\ref{sec:methodology} we
review the tools used in the analysis of the
data. Sect.~\ref{sec:compactness} collects our results and there we
first discuss the threshold of the bar-mode instability, its
persistence, and eventually the onset of higher-mode dynamical
instabilities. Finally, Sect.~\ref{sec:conclusions} contains our
conclusions and the goals of our future research. Hereafter we use a
space like signature $(-,+,+,+)$, with Greek indices running from 0 to
3, Latin indices from 1 to 3 and the standard convention for the
summation over repeated indices.  Unless explicitly stated, all the
quantities are expressed in units in which $c=G=M_\odot=1$.

%-----------------------------------------------------------------
\section{Initial data and numerical evolution method}
\label{sec:initial}
%-----------------------------------------------------------------

Our simulations involve the numerical solution in three spatial
dimensions (3D) of the full set of Einstein equations coupled to that
of a perfect-fluid matter
\begin{equation}
\label{eq:Einstein}
G_{\mu\nu} = 8 \pi T_{\mu\nu} \;,
\end{equation}
where
\begin{equation}
T^{\mu\nu} = \rho \left(1 +\epsilon + \frac{p}{\rho} \right) 
              u^{\mu} u^{\nu} + p g^{\mu\nu} \;,
\end{equation}
and $u^\mu$ is the fluid 4-velocity, $p$ is the fluid pressure,
$\epsilon$ the specific internal energy and $\rho$ the rest-mass
density, so that $e = \rho (1+\epsilon)$ is the energy density in the
rest frame of the fluid. The evolution of the spacetime must be
supplemented by the evolution of the relativistic hydrodynamics
equations: the conservation laws for the energy-momentum tensor
$\nabla_{\mu} T^{\mu\nu}= 0$ and the baryon number $\nabla_{\mu} (\rho
u^{\mu}) = 0$, complemented with an equation of state (EOS) of type $p
=p(\rho,\epsilon)$.

%%%----------------------------------------------------------------
\begin{figure}[t]
\begin{center}
\includegraphics[width=1.0\textwidth,height=0.75\textwidth]{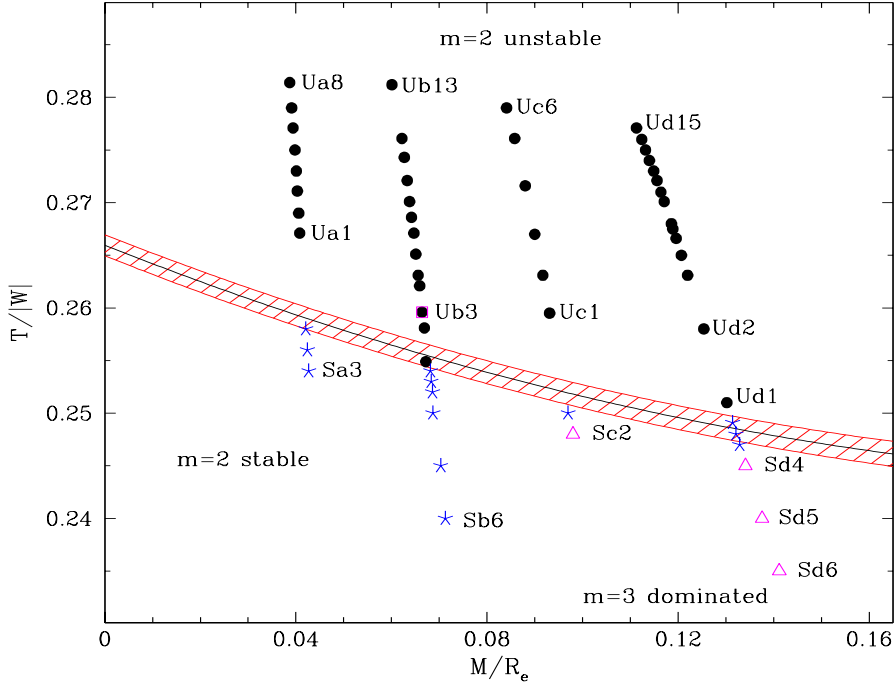}
\end{center}
\vglue-0.8cm
\caption{Position on the $(M/R_e,\beta)$ plane of the considered stellar
  models.  Indicated respectively with stars and filled circles are the
  $m\!\!=\!\!2$-stable and $m\!\!=\!\!2$-unstable models belonging to the
  four sequences of constant rest mass. Triangles refer instead to models
  where the $m\!\!=\!\!3$ deformation is the fastest growing one.
  Indicated with a solid line is the threshold of the bar-mode
  instability, while the dashed region represents the region of our
  estimated error-bars.}
\label{fig:beta vs compactness}
\end{figure}
%%%----------------------------------------------------------------

The initial data for our simulations are computed as stationary
equilibrium solutions for axisymmetric and rapidly rotating
relativistic stars in polar coordinates~\cite{Stergioulas95}. In
generating these equilibrium models the metric describing an
axisymmetric relativistic star is assumed to have the form
\begin{equation} 
\dr s^{2} = - \e^{\mu+\nu} \dr t^{2}
         + \e^{\mu-\nu} r^{2} \sin^{2}\theta (\dr \phi-\omega \dr t)^{2}
         +\e^{2\xi}(\dr r^{2}+r^{2} \dr \theta^{2})\;,
\end{equation} 
where $\mu$, $\nu$, $\omega$ and $\xi$ are space-dependent metric
functions. As in paper I, we assume the matter to be characterized by
a non-uniform angular-velocity distribution of the form
\begin{equation} 
\Omega_{c} -\Omega   =   \frac{r_{e}^{2}}{\hat{A}^{2}}
       \left[ \frac{(\Omega-\omega) r^2   \sin^2\theta \e^{-2\nu}
              }{1-(\Omega-\omega)^{2} r^2 \sin^2\theta \e^{-2\nu}
       }\right] \;,
\label{eq:velocityProfile}
\end{equation} 
where $r_{e}$ is the coordinate equatorial stellar radius and the
coefficient $\hat{A}$ is a measure of the degree of differential
rotation, which we set to $\hat{A}=1$.  All the equilibrium models
considered here have been calculated using the relativistic polytropic
EOS ($p=K\rho^\Gamma$) with $K=100$ and $\Gamma=2$ and are members of
four sequences having a constant rest mass $M_*$ equal to $1.0\ M_{\odot}$,
$1.51\ M_{\odot}$, $2\ M_{\odot}$ and $2.5\ M_{\odot}$,
respectively. The main properties of the four sequences are reported
in tables \ref{table:initialA}-\ref{table:initialD}, which report the
{\it baryonic} mass $M_*$, the gravitational mass $M$, the angular
momentum $J$, the rotational kinetic energy $T$, the gravitational
binding energy $W$ and the instability parameter $\beta = T /|W|$, and
whose definitions are
\begin{eqnarray}
&M_* \equiv \displaystyle \int\!d^3\!x\,  \sqrt{\gamma} W_{_{\!L}} \rho \,, 
\qquad 
&M \equiv \displaystyle \int\!d^3\!x\, \left( -2T^0_0+T^\mu_\mu\right) 
	\alpha \sqrt{\gamma}\,,
\label{eq:DEF M}
\\
&{E_{\rm int}} \equiv \displaystyle \int\!d^3\!x\,  \sqrt{\gamma} W_{_{\!L}}
\rho \epsilon\,, 
\qquad
&J \equiv  \displaystyle \int\!d^3\!x\, T^0_\phi \alpha \sqrt{\gamma}\,,
\label{eq:DEF J}
\\
&T \equiv \displaystyle \frac{1}{2} \int\!d^3\!x\, \Omega  T^0_\phi
\alpha \sqrt{\gamma}\,, 
&W\equiv  T + {E_{\rm int}} + M_* - M \,,
 \label{eq:DEF W}
\end{eqnarray}
where $\alpha$ is the lapse function, $\sqrt{\gamma}$ is the square
root of the three-dimensional metric determinant and $W_{_{\!L}}=\alpha
u^0$ is the fluid Lorentz factor. We stress that that the
definitions~(\ref{eq:DEF M})--(\ref{eq:DEF W}) of quantities such as
$J$, $T$, $W$ and $\beta$ are meaningful only in the case of
stationary axisymmetric configurations and should therefore be treated
with care once the rotational symmetry is lost.

Traditionally, numerical simulations of the dynamical bar-mode
instability have been sometimes sped up by introducing very large
$m\!\!=\!\!2$ deformations in the initial condition. As
discussed in paper I, the introduction of any perturbation (especially
when this is not a small one) may lead to spurious effects and
erroneous interpretations. Being aware of this, we used, only in some
selected simulations below the threshold, initial density
perturbations of the type
\begin{equation}
\label{eq:densPert}
\delta\rho_2(x,y,z) = \delta_{2} 
	\left(\frac{x^{2}-y^{2}}{r_{e}^{2}}\right) \rho \;,
\end{equation}
where $\delta_2$ is the amplitude of the $m\!\!=\!\!2$ perturbation
(which we set to be $\delta_2\simeq 0.01-0.04$). This perturbation has
then the effect of superimposing on the axially symmetric initial
model a bar deformation that is larger than the (unavoidable)
$m\!\!=\!\!4$-mode perturbation introduced by the Cartesian grid
discretization. The introduction of such perturbation allowed us to
estimate the frequency of the $m\!\!=\!\!2$ mode below the threshold
for the onset of the instability and reduce considerably the computing
costs in a region of the parameter space where the instability does
not develop.

%%%----------------------------------------------------------------
%%%----------------------------------------------------------------
\begin{table}[t]
%\vspace{-4.3cm}
%\hspace{-2.5cm}
\newcommand{\msec}{{$\scriptstyle \mathrm{(ms)}$}}
\newcommand{\expq}{{$\scriptstyle (10^{-4})$}}
\caption{Main properties of the stellar models of the sequence with
  $M_*=1M_{\odot}$. Starting from the left: the name of the
  simulation, the compactness $M/R_e$, the instability parameter
  $\beta$, the central rest-mass density $\rho_c$, the ratio between
  the polar and the equatorial coordinate radii $r_{p}/r_{e}$, the
  proper equatorial radius $R_{e}$, the gravitational mass $M$, the
  total angular momentum $J$ divided by the square of the
  gravitational mass, the rotational periods at the axis $P_{a}$ and
  at the equator $P_{e}$. The initial letter in the model's name
  indicates whether it is an unstable (U) or a stable (S)
  configuration.}
\label{table:initialA}
\begin{indented}
\item[]
\hspace{-1.7cm}
\begin{tabular}{@{}cccccccccc}
\br
 Mod &$M/R_e$&$\beta$&$\rho_c$ \expq&
$r_p/r_e$ &$R_e$ &$M $ & $J/M^2$ &$P_a$ \msec &$P_e$ \msec\\[1mm]
\hline
Ua8  & 0.0387 & 0.2814 & 0.4369 & 0.21044 & 25.37 & 0.982& 2.142 & 2.283 & 4.945  \\
Ua7  & 0.0391 & 0.2790 & 0.5662 & 0.23177 & 25.04 & 0.980& 2.098 & 2.207 & 4.791  \\
Ua6  & 0.0394 & 0.2771 & 0.6429 & 0.24288 & 24.83 & 0.979& 2.070 & 2.164 & 4.702  \\
Ua5  & 0.0398 & 0.2750 & 0.7087 & 0.25173 & 24.63 & 0.979& 2.044 & 2.127 & 4.626  \\
Ua4  & 0.0401 & 0.2730 & 0.7672 & 0.25918 & 24.45 & 0.980& 2.021 & 2.094 & 4.559  \\
Ua3  & 0.0403 & 0.2711 & 0.8168 & 0.26550 & 24.29 & 0.978& 2.001 & 2.068 & 4.506  \\
Ua2  & 0.0406 & 0.2690 & 0.8668 & 0.27149 & 24.11 & 0.979& 1.979 & 2.040 & 4.449  \\
Ua1  & 0.0408 & 0.2671 & 0.9107 & 0.27684 & 23.95 & 0.978& 1.960 & 2.017 & 4.402  \\
\hline
Sa1  & 0.0421 & 0.2580 & 1.0907 & 0.29807 & 23.22 & 0.977& 1.878 & 1.922 & 4.209  \\
Sa2  & 0.0424 & 0.2560 & 1.1270 & 0.30236 & 23.05 & 0.977& 1.861 & 1.904 & 4.172  \\
Sa3  & 0.0427 & 0.2540 & 1.1619 & 0.30655 & 22.89 & 0.977& 1.844 & 1.886 & 4.135  \\
\br
\end{tabular}
\end{indented}
\end{table}
%%%%%%%%%%%%%%%%%%%%%%%%%%%%%%%%%%

\begin{table}[t]
%\vspace{-4.3cm}
%\hspace{-2.5cm}
\newcommand{\msec}{{$\scriptstyle \mathrm{(ms)}$}}
\newcommand{\expq}{{$\scriptstyle (10^{-4})$}}
\caption{Same quantities as in \tref{table:initialA} for the sequence 
of models with $M_*=1.51M_{\odot}$.}
\label{table:initialB}
\begin{indented}
\item[]
\hspace{-1.7cm}
\begin{tabular}{@{}cccccccccc}
\br
 Mod &$M/R_e$&$\beta$&$\rho_c$ \expq&
$r_p/r_e$ &$R_e$ &$M $ & $J/M^2$ &$P_a$ \msec &$P_e$ \msec\\[1mm]
\hline
Ub13 & 0.0601 & 0.2812 & 0.5990 & 0.20012 & 24.31 & 1.462& 1.753 & 1.723 & 3.910  \\
Ub12 & 0.0622 & 0.2761 & 0.9938 & 0.24151 & 23.52 & 1.462& 1.679 & 1.599 & 3.655  \\
Ub11 & 0.0626 & 0.2743 & 1.0920 & 0.25012 & 23.31 & 1.460& 1.660 & 1.572 & 3.598  \\
Ub10 & 0.0633 & 0.2721 & 1.1960 & 0.25858 & 23.08 & 1.461& 1.639 & 1.542 & 3.536  \\
Ub9  & 0.0638 & 0.2701 & 1.2844 & 0.26554 & 22.88 & 1.460& 1.621 & 1.517 & 3.486  \\
Ub8  & 0.0642 & 0.2686 & 1.3465 & 0.27028 & 22.73 & 1.460& 1.608 & 1.500 & 3.451  \\
Ub7  & 0.0646 & 0.2671 & 1.4055 & 0.27474 & 22.59 & 1.459& 1.596 & 1.485 & 3.418  \\
Ub6  & 0.0651 & 0.2651 & 1.4812 & 0.28033 & 22.40 & 1.459& 1.579 & 1.465 & 3.377  \\
Ub5  & 0.0656 & 0.2631 & 1.5534 & 0.28560 & 22.22 & 1.459& 1.564 & 1.446 & 3.339  \\
Ub4  & 0.0659 & 0.2621 & 1.5879 & 0.28813 & 22.13 & 1.458& 1.557 & 1.437 & 3.321  \\
Ub3  & 0.0664 & 0.2595 & 1.6730 & 0.29433 & 21.91 & 1.456& 1.539 & 1.416 & 3.278  \\
Ub2  & 0.0669 & 0.2581 & 1.7233 & 0.29779 & 21.78 & 1.457& 1.527 & 1.403 & 3.251  \\
Ub1  & 0.0674 & 0.2551 & 1.8120 & 0.30450 & 21.54 & 1.452& 1.509 & 1.384 & 3.210  \\
 \hline
Sb1  & 0.0682 & 0.2541 & 1.8600 & 0.30691 & 21.42 & 1.461& 1.497 & 1.368 & 3.179  \\
Sb2  & 0.0682 & 0.2530 & 1.8845 & 0.30915 & 21.35 & 1.456& 1.492 & 1.364 & 3.171  \\
Sb3  & 0.0684 & 0.2520 & 1.9155 & 0.31134 & 21.27 & 1.456& 1.485 & 1.357 & 3.156  \\
Sb4  & 0.0687 & 0.2503 & 1.9620 & 0.31500 & 21.14 & 1.452& 1.476 & 1.348 & 3.137  \\
Sb5  & 0.0703 & 0.2451 & 2.1280 & 0.32600 & 20.70 & 1.456& 1.439 & 1.308 & 3.057  \\
Sb6  & 0.0713 & 0.2403 & 2.2610 & 0.33600 & 20.32 & 1.449& 1.411 & 1.282 & 3.002  \\
\br
\end{tabular}
\end{indented}
\end{table}

\begin{table}[t]
%\vspace{-4.3cm}
%\hspace{-2.5cm}
\newcommand{\msec}{{$\scriptstyle \mathrm{(ms)}$}}
\newcommand{\expq}{{$\scriptstyle (10^{-4})$}}
\caption{Same quantities as in \tref{table:initialA} for the sequence 
of models with $M_*=2M_{\odot}$.}
\label{table:initialC}
\begin{indented}
\item[]
\hspace{-1.7cm}
\begin{tabular}{@{}cccccccccc}
\br
 Mod &$M/R_e$&$\beta$&$\rho_c$ \expq&
$r_p/r_e$ &$R_e$ &$M $ & $J/M^2$ &$P_a$ \msec &$P_e$ \msec\\[1mm]
\hline
\hline
Uc6 & 0.0841 & 0.2790 & 0.9669 & 0.21142 & 22.82 & 1.920& 1.495 & 1.317 & 3.161  \\
Uc5 & 0.0858 & 0.2761 & 1.2663 & 0.23377 & 22.32 & 1.916& 1.460 & 1.260 & 3.042  \\
Uc4 & 0.0880 & 0.2716 & 1.6079 & 0.25516 & 21.74 & 1.913& 1.420 & 1.200 & 2.916  \\
Uc3 & 0.0900 & 0.2670 & 1.8982 & 0.27117 & 21.23 & 1.911& 1.384 & 1.152 & 2.815  \\
Uc2 & 0.0917 & 0.2631 & 2.1264 & 0.28298 & 20.82 & 1.908& 1.356 & 1.116 & 2.741  \\
Uc1 & 0.0931 & 0.2595 & 2.3176 & 0.29240 & 20.48 & 1.907& 1.333 & 1.088 & 2.682  \\
\hline
Sc1 & 0.0970 & 0.2500 & 2.8043 & 0.31526 & 19.61 & 1.902& 1.275 & 1.021 & 2.543  \\
Sc2 & 0.0980 & 0.2480 & 2.9091 & 0.31987 & 19.42 & 1.902& 1.263 & 1.007 & 2.514  \\
\br
\end{tabular}
\end{indented}
\end{table}

\begin{table}[t]
%\vspace{-4.3cm}
%\hspace{-2.5cm}
\newcommand{\msec}{{$\scriptstyle \mathrm{(ms)}$}}
\newcommand{\expq}{{$\scriptstyle (10^{-4})$}}
\caption{Same quantities as in \tref{table:initialA} for the sequence 
of models with $M_*=2.5M_{\odot}$.}
\label{table:initialD}
\begin{indented}
\item[]
\hspace{-1.7cm}
\begin{tabular}{@{}cccccccccc}
\br
 Mod &$M/R_e$&$\beta$&$\rho_c$ \expq&
$r_p/r_e$ &$R_e$ &$M $ & $J/M^2$ &$P_a$ \msec &$P_e$ \msec\\[1mm]
\hline
Ud15 & 0.1113 & 0.2771 & 1.3117 & 0.21259 & 21.29 & 2.369& 1.320 & 1.025 & 2.630  \\
Ud14 & 0.1124 & 0.2760 & 1.4608 & 0.22121 & 21.07 & 2.368& 1.307 & 1.005 & 2.588  \\
Ud13 & 0.1132 & 0.2750 & 1.5826 & 0.22785 & 20.90 & 2.366& 1.298 & 0.990 & 2.555  \\
Ud12 & 0.1140 & 0.2740 & 1.6961 & 0.23372 & 20.74 & 2.364& 1.289 & 0.976 & 2.526  \\
Ud11 & 0.1149 & 0.2730 & 1.8103 & 0.23925 & 20.58 & 2.364& 1.280 & 0.962 & 2.496  \\
Ud10 & 0.1156 & 0.2721 & 1.9104 & 0.24399 & 20.44 & 2.362& 1.272 & 0.950 & 2.471  \\
Ud9  & 0.1164 & 0.2710 & 2.0153 & 0.24868 & 20.29 & 2.362& 1.264 & 0.938 & 2.445  \\
Ud8  & 0.1171 & 0.2701 & 2.1102 & 0.25286 & 20.16 & 2.360& 1.256 & 0.927 & 2.423  \\
Ud7  & 0.1186 & 0.2680 & 2.3022 & 0.26083 & 19.89 & 2.359& 1.241 & 0.906 & 2.377  \\
Ud6  & 0.1189 & 0.2675 & 2.3534 & 0.26291 & 19.82 & 2.358& 1.238 & 0.901 & 2.366  \\
Ud5  & 0.1196 & 0.2666 & 2.4336 & 0.26609 & 19.71 & 2.357& 1.231 & 0.892 & 2.348  \\
Ud4  & 0.1207 & 0.2650 & 2.5698 & 0.27128 & 19.52 & 2.356& 1.221 & 0.878 & 2.318  \\
Ud3  & 0.1220 & 0.2631 & 2.7402 & 0.27760 & 19.29 & 2.353& 1.208 & 0.861 & 2.282  \\
Ud2  & 0.1254 & 0.2580 & 3.1583 & 0.29210 & 18.74 & 2.349& 1.178 & 0.821 & 2.198  \\
Ud1  & 0.1302 & 0.2510 & 3.7335 & 0.31027 & 18.00 & 2.343& 1.138 & 0.772 & 2.094  \\
\hline
Sd1  & 0.1314 & 0.2491 & 3.8899 & 0.31503 & 17.81 & 2.341& 1.127 & 0.760 & 2.068  \\
Sd2  & 0.1321 & 0.2480 & 3.9735 & 0.31745 & 17.71 & 2.340& 1.122 & 0.753 & 2.054  \\
Sd3  & 0.1329 & 0.2470 & 4.0607 & 0.31991 & 17.61 & 2.340& 1.116 & 0.746 & 2.040  \\
Sd4  & 0.1341 & 0.2450 & 4.2188 & 0.32450 & 17.42 & 2.337& 1.106 & 0.735 & 2.016  \\
Sd5  & 0.1376 & 0.2400 & 4.6380 & 0.33596 & 16.95 & 2.333& 1.081 & 0.706 & 1.955  \\
Sd6  & 0.1412 & 0.2350 & 5.0710 & 0.34714 & 16.49 & 2.329& 1.056 & 0.678 & 1.897  \\
\br
\end{tabular}
\end{indented}
\end{table}
%%%----------------------------------------------------------------
%%%----------------------------------------------------------------

We solve the Einstein equations \eref{eq:Einstein} formulated as a
first-order (in time) quasi-linear~\cite{Richtmyer67} system of
equations, where the independent variables are the three-metric
$\gamma_{ij}$ and the extrinsic curvature $K_{ij}$. In particular, we
use the conformal traceless reformulation of the ADM system of
evolution equations, first suggested in ref.~\cite{Nakamura:87}, in
which the evolved variables are the conformal factor $\phi$, the trace
of the extrinsic curvature $K$ the conformal 3-metric
$\tilde{\gamma}_{ij}$, the conformal traceless extrinsic curvature
$\tilde{A}_{ij}$ and the {\it conformal connection functions }
$\tilde{\Gamma}^i$. The solution of the hydrodynamics equations is
obtained by using the general-relativistic hydrodynamics code {\tt
Whisky}, in which the hydrodynamics equations are written as finite
differences on a Cartesian grid and solved using high-resolution
shock-capturing schemes, as described in ref.~\cite{Baiotti03a}.
During the evolution we use the ``ideal-fluid'' EOS: $p=(\Gamma-1)
\rho \epsilon$.  Full details of the numerical scheme and the gauge
conditions used are reported in paper I.

%XXXXXXXXXXXXXXXXXXXXXXXXXXXXXXXXXXXXXXXXXXXXXXXXXXXXXXXXXXXXXXXXXXXXXXXXXX%
%-----------------------------------------------------------------
\section{Methodology of the analysis}
\label{sec:methodology}
%-----------------------------------------------------------------
%XXXXXXXXXXXXXXXXXXXXXXXXXXXXXXXXXXXXXXXXXXXXXXXXXXXXXXXXXXXXXXXXXXXXXXXXXX%

A number of different quantities are calculated during the evolution
to monitor the dynamics of the instability. Among them is the
quadrupole moment of the matter distribution, which we compute in
terms of the conserved density $\sqrt{\gamma} W_{_{\!L}} \rho$ rather
than of the rest-mass density $\rho$ or of the $T_{00}$ component of
the stress energy momentum tensor
\begin{equation}
\label{eq:defQuadrupole}
I^{jk} = \int\! d^{3}\!x \; \sqrt{\gamma} W_{_{\!L}} \rho  \; x^{j} x^{k} \;.
\end{equation}
Of course, the use of $\sqrt{\gamma} W_{_{\!L}} \rho$ in place of
$\rho$ or of $T_{00}$ is arbitrary and all the three expressions would
have the same Newtonian limit. However, we prefer the form
(\ref{eq:defQuadrupole}) because $\sqrt{\gamma}W_{_{\!L}}\rho$ is a
quantity whose conservation is guaranteed by the form chosen for the
hydrodynamics equations.  The quantity (\ref{eq:defQuadrupole}) can be
conveniently used to quantify both the growth time of the instability
$\tau_2$ and the oscillation frequency of the unstable bar-mode once
the instability is fully developed $f_2$. (Hereafter we will indicate
respectively with $\tau_i$ and $f_i$ the growth time and frequency of
the $m=i$ unstable mode.)

In practice, we perform a nonlinear least-square fit of the computed
quadrupole $I^{jk}(t)$ and we generally use the $xy$ component, with
the trial functions
\begin{equation}  
I^{jk}(t) = (I^{jk})_{0} \; \e^{t/\tau_2} 
               \cos(2\pi\, f_2 \, t+\phi_{0})\;.
\label{eq:Quadrupolefit}
\end{equation}  
Furthermore, we define the modulus $I(t)$ of the two components of the
quadrupole in the $xy$ plane and the distortion parameter $\eta(t)$
as
\begin{equation}
\label{eq:modQuadrupole}
I \equiv \frac{1}{2}\sqrt{(2I^{xy})^2+(I^{xx}-I^{yy})^2}\;, 
\quad 
\label{eq:eta}
\eta \equiv \frac{I}{2(I^{xx}+I^{yy})}  \;. 
\end{equation}
and the instantaneous orientation of the bar is given by 
\begin{equation}
\label{eq:phyQuadrupole}
\phi_{\rm bar} = \tan^{-1}\left(\frac{2I^{xy}}{I^{xx}-I^{yy}}\right)\;. 
\end{equation}

Finally,  as a useful tool to describe the nonlinear properties of 
the development and saturation of the instability, the rest-mass density 
is decomposed into its Fourier modes $P_m(t)$:
\begin{equation}
{P}_m \equiv\int\!d^3\!x  \, \rho \, \e^{{\rm i} m \phi}\;.
%% \quad {\phi}_m  \, . 
\label{eq:modes}
\label{eq:phimodes}
\end{equation}
The phase ${\phi}_m\equiv \arg ({P}_m)$ essentially provides the
instantaneous orientation of the $m$-th mode when the corresponding
mode has a nonzero power.  Note that despite their denomination, the
Fourier modes (\ref{eq:modes}) do not represent proper eigenmodes of
oscillation of the star. While, in fact, the latter are well defined
only within a perturbative regime, the former simply represent a tool
to quantify, within the fully nonlinear regime, what are the main
components of the rest-mass distribution. Stated differently, we do
not expect quasi-normal modes of oscillations to be present but in the
initial and final stages of the instability, for which a perturbative
description is adequate.

While all quantities (\ref{eq:defQuadrupole})--(\ref{eq:phimodes}) are
expressed in terms of the coordinate time $t$ and do not represent
therefore invariant measurements, the lengthscale of variation of the
lapse function at any given time is always larger than twice the
stellar radius at that time, ensuring that events on the same
timeslice are also close in proper time.  As representative
examples, we note that for the most compact model the values of the
lapse at the centre of the star, at its surface and at the outer
boundary are $0.67, 0.84$ and $0.95$, respectively. Similarly, the
corresponding values for the least compact model are $0.92, 0.95$ and
$0.98$, respectively.

The simulations have been carried out on a grid with a uniform
resolution of $\Delta x/M_\odot=0.625$ and outer boundaries at
$48.75\,M_{\odot}$, where ``radiative'' boundary conditions
(\textit{i.e.}, Sommerfeld outgoing boundary conditions) are applied
to the field variables, while the fluid variables are simply not
evolved.  Such outer boundaries are sufficiently far from the surface
of the star to make the use of mesh refinements in {\tt Whisky} not
necessary (see also Section VIE of paper I for a more detailed
discussion of the role of the grid size on the evolution of the
instability). Furthermore, for those models used in the extrapolation
technique and that are largely over-critical (see
Sect.~\ref{sec:threshold}), we have imposed a ``bitant'' symmetry
(\textit{i.e.,} $z\rightarrow -z$) and a ``$\pi$-symmetry''
(\textit{i.e.}, a $180^o$ rotation around the $z$-axis) in order to
reduce the size of the computational domain of a factor of 4.

%%%%%%%%%%%%%%%%%%%%%%%%%%%%%%%%%%%%%%%%%%%%%%%%%%%%%%%%%%%%%%%%%%%%%%%%
%%
%%%%%%%%%%%%%%%%%%%%%%%%%%%%%%%%%%%%%%%%%%%%%%%%%%%%%%%%%%%%%%%%%%%%%%%%
\begin{table}
\caption{\label{table:fit_5_35_beta} Least-square fit of the value of
  $\beta$ at the threshold for the development of the bar-mode
  instability for the four series of models reported in
  \tref{table:unstable}. The critical value for the onset of the
  instability $\beta_c$ is the value of $\beta$ for $1/\tau_2^2=0$
  ($\tau$ is measured in ms) and the digits in brackets represent the
  error in the fit. The results of the fits are shown in
  \fref{fig:NEW_EXTRAP}.}
\begin{indented}
\item[]
\begin{tabular}{@{}rlll}
\br
$M_*=1.0\,M_{\odot}$& $\beta=0.2598(8)  $&$ + 0.0379(19) $&$ (1/\tau_2)^2  $\\
$M_*=1.5\,M_{\odot}$& $\beta=0.2558(5)  $&$ + 0.0236(8)  $&$ (1/\tau_2)^2  $\\
$M_*=2.0\,M_{\odot}$& $\beta=0.2528(15) $&$ + 0.0161(13) $&$ (1/\tau_2)^2  $\\
$M_*=2.5\,M_{\odot}$& $\beta=0.2494(14) $&$ + 0.0116(8)  $&$ (1/\tau_2)^2  $\\
\br
\end{tabular}
\end{indented}
\end{table}
%%%%%%%%%%%%%%%%%%%%%%%%%%%%%%%%%%%%%%%%%%%%%%%%%%%%%%%%%%%%%%%%%%%%%%%%
%%
%%%%%%%%%%%%%%%%%%%%%%%%%%%%%%%%%%%%%%%%%%%%%%%%%%%%%%%%%%%%%%%%%%%%%%%%

%%%%%%%%%%%%%%%%%%%%%%%%%%%%%%%%%%%%%%%%%%%%%%%%%%%%%%%%%%%%%%%%%%%%%%%%
%%
%%%%%%%%%%%%%%%%%%%%%%%%%%%%%%%%%%%%%%%%%%%%%%%%%%%%%%%%%%%%%%%%%%%%%%%%
\begin{table}
\caption{\label{table:fit_5_35_freq} Least-square fit of the value of
  the frequency $f_{\mathrm B}$ (in Hz) of the bar-mode
  (\Eref{eq:Quadrupolefit}) at the threshold for the onset of the
  bar-mode instability as a function of $\theta \equiv
  (\beta-\beta_c)/\beta_c$ for the four series of models at constant
  baryonic mass. The value of the frequency $f_{2}$ at the threshold
  is the value for $\theta=0$ and the digits in brackets represent the
  error in the fit.}
\begin{indented}
\item[]
\begin{tabular}{@{}rlll}
\br
$M_*=1.0\,M_{\odot}$& $f_{\mathrm B}=   384(    7)  $ & $-8(25)\, 10 \, \theta$ & $ -6(2)\, 1000\, \theta^2 $\\
$M_*=1.5\,M_{\odot}$& $f_{\mathrm B}=   551(    8)  $ & $-4(3) \, 100\, \theta$ & $ -6(2)\, 1000\, \theta^2 $\\
$M_*=2.0\,M_{\odot}$& $f_{\mathrm B}=   738(   32)  $ & $-7(11)\, 100\, \theta$ & $ -6(8)\, 1000\, \theta^2 $\\
$M_*=2.5\,M_{\odot}$& $f_{\mathrm B}=   991(   10)  $ & $-10(3)\, 100\, \theta$ & $ -9(2)\, 1000\, \theta^2 $\\
\br
%$M_*=1.0\,M_{\odot}$& $f_{\mathrm B}=   384(    7)  $ & $   -82(  256)\theta  $ & $ -5979( 2288)\theta^2 $\\
%$M_*=1.5\,M_{\odot}$& $f_{\mathrm B}=   551(    8)  $ & $  -393(  277)\theta  $ & $ -6138( 2261)\theta^2 $\\
%$M_*=2.0\,M_{\odot}$& $f_{\mathrm B}=   738(   32)  $ & $  -728( 1096)\theta  $ & $ -6300( 8282)\theta^2 $\\
%$M_*=2.5\,M_{\odot}$& $f_{\mathrm B}=   991(   10)  $ & $ -1030(  283)\theta  $ & $ -8719( 1844)\theta^2 $\\
%\br
\end{tabular}
\end{indented}
\end{table}
%%%%%%%%%%%%%%%%%%%%%%%%%%%%%%%%%%%%%%%%%%%%%%%%%%%%%%%%%%%%%%%%%%%%%%%%
%%
%%%%%%%%%%%%%%%%%%%%%%%%%%%%%%%%%%%%%%%%%%%%%%%%%%%%%%%%%%%%%%%%%%%%%%%%

%-----------------------------------------------------------------
\section{Effects of the compactness}
\label{sec:compactness}
%-----------------------------------------------------------------

%%%----------------------------------------------------------------
%%% Fig: FREQ FIT
%%%----------------------------------------------------------------
%\includegraphics[width=16truecm,height=16truecm]{\imgname{freqfit}}
%\hskip 0.5cm
%\includegraphics[width=7.5truecm]{\imgname{critical_fit}}
\begin{figure}[tbf]
\begin{center}
\hspace{0.09\textwidth}
\includegraphics[width=0.89\textwidth]{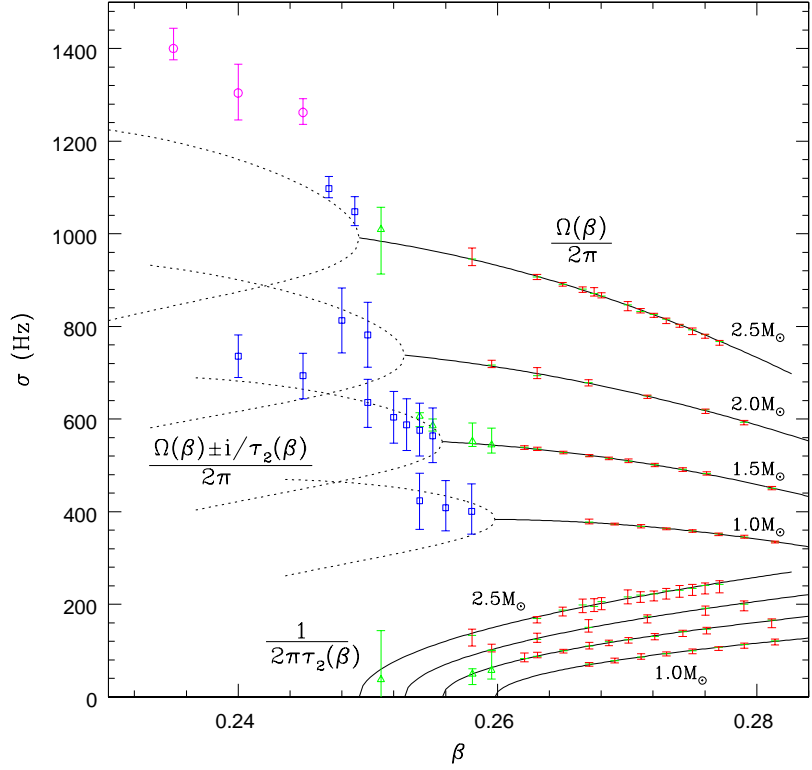}
\end{center}
\vglue-1.2cm
\caption{Critical diagram as constructed with the frequencies and growth
  times relative to the unperturbed models of \tref{table:unstable}. The
  solid lines represent the two fitted curves for $\Omega(\beta)$ and
  $\tau_2(\beta)$, while the dotted lines the corresponding
  extrapolations below the threshold. Triangles refer to the unperturbed
  models of \tref{table:unstable} that have not been used for the fit,
  squares to the perturbed models of \tref{table:stable} (squares), and
  open circles to the the models dominated by the $m\!\!=\!\!3$
  deformation.}
\label{fig:freqfit} \label{fig:critical_fit}
\end{figure}
%%%----------------------------------------------------------------
%%% END
%%%----------------------------------------------------------------

%-----------------------------------------------------------------
\subsection{Threshold of the $m\!\!=\!\!2$ instability}
\label{sec:threshold}
%-----------------------------------------------------------------

The determination of the dependence on $\beta$ of the frequencies and
of the growth times of the $m\!\!=\!\!2$ bar-mode instability in the
region near the threshold is particularly delicate as the models are
only slightly over-critical, with very small growth rates and hence
the simulations are computationally very expensive. For this purpose
we here use an extrapolation technique already described in paper I,
where it was shown to be both accurate and robust. In essence, we
exploit the results of the classical Newtonian study of the bar-mode
instability of Maclaurin spheroids of incompressible and
self-gravitating Newtonian fluid in
equilibrium~\cite{Chandrasekhar69}, extrapolating, via suitable fits,
its predictions to a general-relativistic context. We recall, in fact,
that in the classical scenario the eigenfrequency of the $m\!\!=\!\!2$
bar-mode can be expressed in terms of two real and differentiable
functions of $\beta$: $\Omega \equiv 2\pi f_2$ and $1/\tau_2^2$, in a
relation of the type
\begin{equation}
\sigma =\Omega(\beta)\pm 
	{\frac{{\mathrm i}}{\sqrt{\tau_2^2(\beta)}}}\;.
\label{eq:CHANDRAbis}
\end{equation}
The bar-mode becomes unstable when the function $1/\tau_2^2$ changes
sign, with the square root going from being imaginary to being real.
The value of $\beta$ at which this change of sign happens represents
then the threshold for the onset of the instability $\beta_c$;
clearly, for models above the threshold, $\Omega/4\pi$ and $\tau_2$
represent the pattern speed and the growth time of the unstable bar
deformation of the considered star model, respectively.

%%%%%%%%%%%%%%%%%%%%%%%%%%%%%%%%%%%%%%%%%%%%%%%%%%%%%%%%%%%%%%%%%%%%%%%
%%%%%%%%%%%%%%%%%%%%%%%%%%%%%%%%%%%%%%%%%%%%%%%%%%%%%%%%%%%%%%%%%%%%%%%
\begin{table}
\newcommand{\SKIP}{{~~~~}}
\newcommand{\comp}{{$M/R_e$}}
\newcommand{\rhoc}{{$\rho_{c}$}}
\newcommand{\Rrate}{{$r_{p}/r_{e}$}}
\newcommand{\msec}{{\small (ms)}}
\newcommand{\expq}{~~{$\scriptstyle (10^{-4})$}~~}
\newcommand{\expD}{~~{$\scriptstyle (10^{-2})$}~~}
\newcommand{\CCS}{\multicolumn{1}{c}{*}}

\caption{For models below the threshold for the onset of the $m \!\! = 
  \!\! 2$ bar-mode instability the table reports the measured frequencies $f_2$ 
  of the bar deformation as well as the frequencies 
  $f_{3}$ and growth times $\tau_3$ of the $m\!\!=\!\!3$ deformation.
  For each model we also indicate the value of $\beta$, the grid
  resolution $\Delta x/M_{\odot}$ and the initial bar-mode perturbation
  $\delta_2$. All models have been evolved  without a  $\pi$-symmetry.}
\label{table:stable}
\begin{indented}
\item[]
\begin{tabular}{@{}lccccrr}
\br
Model& $\beta$ & $\Delta x/M_\odot$ &$\delta_2$&
\multicolumn{1}{c}{$f_2$ (Hz)} & 
\multicolumn{1}{c}{$f_{3}$ (Hz)}       & 
\multicolumn{1}{c}{$\tau_3$ (ms)}      \\
\mr %%%%%%%%%%%%%%%%%%%%%%%%%%%& %%%%%%%%%%%%%%%%%%%%%%%%%%%%%%%%
Sa3   & 0.254  & 0.5 &$0.30$ & $424_{-62}^{+59}$     & \CCS& \CCS\\
Sa2   & 0.256  & 0.5 &$0.01$ & $409_{-50}^{+58}$     & \CCS& \CCS\\
Sa1   & 0.258  & 0.5 &$0.01$ & $401_{-49}^{+59}$     & \CCS& \CCS\\
\br  %%%%%%%&&%0.625%&%%%%%%%%%%%%%%%%%%%
Sb6   & 0.240  & 0.5 &$0.04$ & $736_{-46  }^{+46} $   & \CCS&\CCS \\
Sb5   & 0.245  & 0.5 &$0.04$ & $694_{-50  }^{+48} $   & \CCS&\CCS \\
Sb4   & 0.250  & 0.5 &$0.04$ & $636_{-54  }^{+50}$    & \CCS&\CCS \\
Sb3   & 0.252  & 0.5 &$0.04$ & $604_{-56  }^{+56}$    & \CCS&\CCS \\
Sb2   & 0.253  & 0.5 &$0.04$ & $588_{-56  }^{+56}$    & \CCS&\CCS \\
Sb1   & 0.254  & 0.5 &$0.04$ & $576_{-56  }^{+58}$    & \CCS&\CCS \\
Ub1   & 0.255  & 0.5 &$0.04$ & $564_{-58  }^{+60}$    & \CCS&\CCS \\
\hline %%%%%%%%& 0.5&&&&&&&&&&&&&&&&&&&&&&&%%%%%%%%%
Sb1   & 0.254  & 0.5 &$0.0$  & $606_{-21 }^{+8}$     & $1302^{-12}_{+12}$ & 3.8 \\ % time 10-75%
Ub1   & 0.255  & 0.5 &$0.0$  & $586_{-12  }^{+14 }$  & $1284^{-12}_{+16}$ & 10.0\\
\br  %%%%%%%&&%0.625%&%%%%%%%%%%%%%%%%%%%
Sc1   & 0.250  & 0.625 &$0.01$   & $782_{-70  }^{+70}$  & \CCS & \CCS \\ %  time 5-75%
Sc2   & 0.248  & 0.625 &$0.01$   & $818_{-70  }^{+70}$  & \CCS & \CCS \\ %  time 5-75%
Sc2   & 0.248  & 0.625 &$0.0$    & $\star$  & $1746^{+12}_{-14}$ & 4.0\\ %  time 5-75%
\br %%%%%%%%& %&%%%%%%%%%%%%%%%%%%%%%%%%%%%%   
Sd6   & 0.235  & 0.625 &$0.01$ & $1400_{-24  }^{+44} $  & $2454^{+52}_{-58}$  & 2.7 \\ %102 193x193x68b pert01 %time 30-75%
Sd5   & 0.240  & 0.625 &$0.01$ & $1304_{-58  }^{+62 }$  & $2312^{+122}_{-100}$& 1.4 \\ %103 193x193x68b pert01
Sd4   & 0.245  & 0.625 &$0.01$ & $1262_{-26  }^{+30 } $ & $2248^{+66}_{-76}$  & 2.9 \\ %104 193x193x68b pert01
Sd3   & 0.247  & 0.625 &$0.01$ & $1098_{-20  }^{+26 }$  & $2200^{+62}_{-54}$  & 2.5 \\ %105 193x193x68b pert01
Sd1   & 0.249  & 0.625 &$0.01$ & $1048_{-30  }^{+32 }$  & $2136^{+74}_{-50}$  & 3.5 \\ %106 193x193x68b pert01
\hline %%%%%%%&& 0.625%&%%%%%%%%%%%%%%%%%%%%%          
Sd6   & 0.235  & 0.625 &$0.0$    & $\star$  & $2474^{+18}_{-20}$ & 2.7\\  % time 5-75%
Sd5   & 0.240  & 0.625 &$0.0$    & $\star$  & $2372^{+20}_{-18}$ & 2.7\\
Sd4   & 0.245  & 0.625 &$0.0$    & $\star$  & $2296^{+16}_{-16}$ & 2.2\\
Sd2   & 0.248  & 0.625 &$0.0$    & $\star$  & $2228^{+22}_{-20}$ & 2.5\\   
Sd1   & 0.249  & 0.625 &$0.0$    &  $ 1446^{+32}_{-26} $  & $2174^{+28}_{-28}$ &2.2\\ %
Ud1   & 0.251  & 0.625 &$0.0$    &  $ 1022^{+32}_{-56} $  & $2158^{+12}_{-16}$ &2.9\\ %
%%%%%%%&%%%%%%%%%%%%%%%%%%%%%
\br
\end{tabular}\\
$*$ A dynamical $m\!\!=\!\!3$ deformation instability cannot be detected.\\
$\star$ The frequency of the $m\!\!=\!\!2$ bar-mode cannot be measured.
\end{indented}
\end{table}
%%%\end{center}
%%% ------------------------------------------------------------------

With the rather reasonable assumption that the two functions $\Omega$
and $1/\tau_2^2$ are regular also in full General Relativity, we
expand them in a Taylor series around the threshold and expressed them
in terms of five unknown coefficients $f_c, f^{(1)}_c, f^{(2)}_c, k,
\beta_c$, \textit{i.e.}
\begin{eqnarray}
&&\frac{\Omega(\beta)}{2\pi} 
       \approx f_c+f^{(1)}_c\frac{(\beta-\beta_c)}{\beta_c}+ 
	f^{(2)}_c\frac{(\beta-\beta_c)^2}{\beta_c^2} + 
        {\cal O}((\beta-\beta_c)^3) \,,\label{eq:fit Omega}
\\
&&\frac{1}{\tau_2^2} \approx \frac{1}{k^2} (\beta-\beta_c) 
	+ {\cal O}((\beta-\beta_c)^2) 
\,. \label{eq:fit beta}
\end{eqnarray}

Expressions~(\ref{eq:fit Omega}) and~(\ref{eq:fit beta}) represent
very good approximations to the actual data and the five parameters
can be determined straightforwardly by fitting the pattern speeds and
the growth times obtained in the largely over-critical models.  For
these models, we recall, the development of the $m\!\!=\!\!2$ bar-mode
deformation is very rapid, the extraction of the instability parameter
is robust and it can be safely simulated even at rather low
resolutions (see paper I).

In practice, using the data obtained from the simulations of the four
sequences of initial models with constant baryonic mass, we have
computed the values for $f_2$ and $\tau_2$ by means of a nonlinear
least-square fit to the trial form of
eq.~(\ref{eq:Quadrupolefit}). Making use of these results, which are
collected in \tref{table:unstable}, we have then computed the unknown
coefficients $f_c$, $f^{(1)}_c$, $f^{(2)}_c$, $k$ and $\beta_c$
through a least-square fit of the values for $f_2$ and $\tau_2$ once
expressed as functions of $\beta$ as in eqs.~(\ref{eq:fit Omega}) and
(\ref{eq:fit beta}).

The results of these fits are reported in
Tables~\ref{table:fit_5_35_beta} and~\ref{table:fit_5_35_freq} and
summarised in Fig.~\ref{fig:critical_fit}. In particular, for each of
the four sequences this figure shows with solid lines the two fitted
curves for $\Omega(\beta)$ and $\tau_2(\beta)$, and with dotted lines
the corresponding extrapolations for models below the threshold.  In
addition, different symbols are used to mark the results of the
numerical simulations, with ``bare'' error bars denoting the
unperturbed unstable models (as reported in Table~\ref{table:unstable}
and which have been used for the fits), triangles denoting the
unperturbed models (as reported in Table~\ref{table:unstable} and not
used for the fits because of the large error in determining their
evolution parameters), squares denoting the stable perturbed models
(as reported in Table~\ref{table:stable}) and open circles denoting
the models dominated by the $m\!\!=\!\!3$ instability (and again
reported in Table~\ref{table:stable}). It is worth noting that for
these last models the frequency of the $m\!\!=\!\!2$ mode is
considerably altered by the growing $m\!\!=\!\!3$ deformation, which
rapidly dominates the small initial $m\!\!=\!\!2$ bar-mode perturbation used
for the simulation of these models.

%%%--------------------------------------------------------------------------
\begin{table}[t]
  \caption{\label{table:unstable} For models above the threshold the
    table reports the measured frequencies $f_2$ 
    and growth times $\tau_2$ of the $m\!\!=\!\!2$ deformation as well as
    the value of $\beta$. All models have been evolved with a grid
    resolution $\Delta x/M_\odot=0.625$ and no initial perturbation;
    furthermore a $\pi$-symmetry has been used for all models with the exception of
    models Ub1 and Ud1 that are closer to the threshold.}
\newcommand{\SKIP}{{~~~~~}}
\newcommand{\comp}{{$M/R_e$}}
\newcommand{\rhoc}{{$\rho_{c}$}}
\newcommand{\Rrate}{{$r_{p}/r_{e}$}}
\newcommand{\msec}{{\small (ms)}}
\newcommand{\expq}{~~{$\scriptstyle (10^{-4})$}~~}
\newcommand{\expD}{~~{$\scriptstyle (10^{-2})$}~~}
\newcommand{\CCS}{\multicolumn{1}{c}{*}}
\begin{indented}
\item[]
\hspace{-0.9cm}
\begin{tabular}{@{}lcllclcll}
%%%%%%%%%%%%%%%%%\hline
\br
%%\cline{1-4}  \cline{6-9}
 Model & $\beta$  &\multicolumn{1}{c}{${\tau}_2$ (ms)} &
 \multicolumn{1}{c}{$f_2$ (Hz)} &&  
 Model & $\beta$  &\multicolumn{1}{c}{${\tau}_2$ (ms)} &
 \multicolumn{1}{c}{$f_2$ (Hz)}\\
\cline{1-4}  \cline{6-9}
%%%%%%%%%%%%%%%%%\hline
Ua1 & 0.2671  &  $2.235_{-0.09}^{+0.16}$ & $377_{-4}^{+7}$  && Uc1 & 0.2596&  $1.564_{-0.17}^{+0.07}$ & $717_{-6}^{+10}$ \\
Ua2 & 0.2690  &  $2.019_{-0.13}^{+0.14}$ & $373_{-1}^{+3}$  && Uc2 & 0.2631&  $1.277_{-0.12}^{+0.07}$ & $694_{-6}^{+18}$ \\
Ua3 & 0.2711  &  $1.838_{-0.14}^{+0.11}$ & $368_{-3}^{+4}$  && Uc3 & 0.2670&  $1.055_{-0.10}^{+0.08}$ & $679_{-7}^{+7}$ \\ 
Ua4 & 0.2730  &  $1.702_{-0.06}^{+0.10}$ & $364_{-3}^{+1}$  && Uc4 & 0.2716&  $0.915_{-0.02}^{+0.08}$ & $650_{-5}^{+2}$ \\ 
Ua5 & 0.2750  &  $1.600_{-0.06}^{+0.10}$ & $359_{-4}^{+2}$  && Uc5 & 0.2761&  $0.829_{-0.02}^{+0.07}$ & $621_{-8}^{+1}$ \\ 
Ua6 & 0.2771  &  $1.482_{-0.02}^{+0.11}$ & $352_{-4}^{+1}$  && Uc6 & 0.2790&  $0.795_{-0.03}^{+0.06}$ & $593_{-6}^{+4}$ \\ 
\cline{6-9}         
Ua7 & 0.2790  &  $1.405_{-0.04}^{+0.11}$ & $345_{-2}^{+4}$  && Ud1 &
0.2510& $\ \ \ \ \approx 4.2$  & $1009_{-96}^{+48}$ \\
%%%%%%%%%%%%%%%%%%%%%%%%4.231_{-3.12}  
Ua8 & 0.2814  &  $1.313_{-0.04}^{+0.10}$ & $335_{-2}^{+2}$  && Ud2 & 0.2580&  $1.193_{-0.11}^{+0.25}$ & $945_{-14}^{+24}$ \\
\cline{1-4}
Ub1 & 0.2551  &  \CCS                  & $586_{-12}^{+14}$&& Ud3 & 0.2631&  $0.938_{-0.02}^{+0.05}$ & $907_{-6}^{+6}$ \\   
Ub2 & 0.2581  &  $3.215_{-0.61}^{+2.62}$ & $552_{-11}^{+40}$&& Ud4 & 0.2650&  $0.846_{-0.03}^{+0.06}$ & $891_{-4}^{+4}$ \\
Ub3 & 0.2595  &  $2.758_{-1.17}^{+1.32}$ & $544_{-17}^{+37}$&& Ud5 & 0.2666&  $0.800_{-0.05}^{+0.07}$ & $881_{-7}^{+5}$ \\ 
Ub4 & 0.2621  &  $1.920_{-0.25}^{+0.17}$ & $540_{-5}^{+3}$  && Ud6 & 0.2675&  $0.813_{-0.06}^{+0.05}$ & $871_{-6}^{+13}$ \\ 
Ub5 & 0.2631  &  $1.803_{-0.15}^{+0.07}$ & $535_{-3}^{+4}$  && Ud7 & 0.2680&  $0.774_{-0.03}^{+0.07}$ & $865_{-4}^{+7}$ \\   
Ub6 & 0.2651  &  $1.591_{-0.04}^{+0.08}$ & $527_{-2}^{+4}$  && Ud8 & 0.2701&  $0.735_{-0.05}^{+0.05}$ & $847_{-13}^{+7}$ \\
Ub7 & 0.2671  &  $1.441_{-0.09}^{+0.10}$ & $520_{-1}^{+4}$  && Ud9 & 0.2710&  $0.724_{-0.02}^{+0.05}$ & $832_{-3}^{+6}$ \\
Ub8 & 0.2686  &  $1.335_{-0.03}^{+0.10}$ & $516_{-3}^{+2}$  && Ud10& 0.2721&  $0.716_{-0.02}^{+0.05}$ & $826_{-7}^{+2}$ \\
Ub9 & 0.2701  &  $1.290_{-0.05}^{+0.07}$ & $509_{-3}^{+5}$  && Ud11& 0.2730&  $0.697_{-0.02}^{+0.06}$ & $816_{-9}^{+3}$ \\
Ub10& 0.2721  &  $1.214_{-0.05}^{+0.08}$ & $502_{-4}^{+3}$  && Ud12& 0.2741&  $0.690_{-0.03}^{+0.05}$ & $804_{-6}^{+1}$ \\
Ub11& 0.2743  &  $1.126_{-0.02}^{+0.09}$ & $492_{-5}^{+2}$  && Ud13& 0.2750&  $0.679_{-0.02}^{+0.05}$ & $793_{-11}^{+4}$ \\
Ub12& 0.2761  &  $1.087_{-0.03}^{+0.08}$ & $483_{-3}^{+4}$  && Ud14& 0.2760&  $0.660_{-0.01}^{+0.06}$ & $782_{-8}^{+1}$ \\
Ub13& 0.2812  &  $0.959_{-0.01}^{+0.10}$ & $450_{-3}^{+4}$  && Ud15& 0.2771&  $0.654_{-0.02}^{+0.05}$ &$767_{-8}^{+2}$ \\
%%%\cline{1-4}  \cline{6-9}
\br\end{tabular}
\\
$\star$ The growth rate of the $m\!\!=\!\!2$ bar-mode cannot be
reliably measured for this model.\\
\end{indented}
\end{table}
%%%%%%%%%%%%%%%%%%%%%%%%%%%%%%%%%%%%%%%%%%%%%%%%%%%%%%%%%%%%%%%%%%%%%%%%
%%
%%%%%%%%%%%%%%%%%%%%%%%%%%%%%%%%%%%%%%%%%%%%%%%%%%%%%%%%%%%%%%%%%%%%%%%%

We also note that the error bars in Table~\ref{table:unstable} are
computed in different ways for the growth rates and for the
frequencies. In the first case they are computed as the difference
between the minimum and maximum values of $d\log(I(t))/dt$ in the time
intervals in which the quadrupolar deformation lies between the 5\% and the
35\% of its maximum amplitude. In the second case, instead, the error
bars are determined using the minimum and maximum values in the time
interval over which the pattern speeds are extracted from the
collective phase $\phi_{\rm bar}(t)$ [\textit{cf.}
  eq.~(\ref{eq:phyQuadrupole})].  The frequencies and the error bars
for the stable models reported in Table~\ref{table:stable} and
Fig.~\ref{fig:critical_fit}, on the other hand, are computed using the
Lomb's power spectrum analysis \cite{lomb76}, which is better suited to
study a signal spanning over a short time interval and which is
comparable with the main frequency of the Fourier transform. In
particular, they refer to the frequency at which the Fourier transform
of the imaginary part of $P_m(t)$ has its maximum, while the error
bars to the interval in frequencies where the Fourier transform is
above $1/2$ of its maximum value.

\begin{figure}[t]
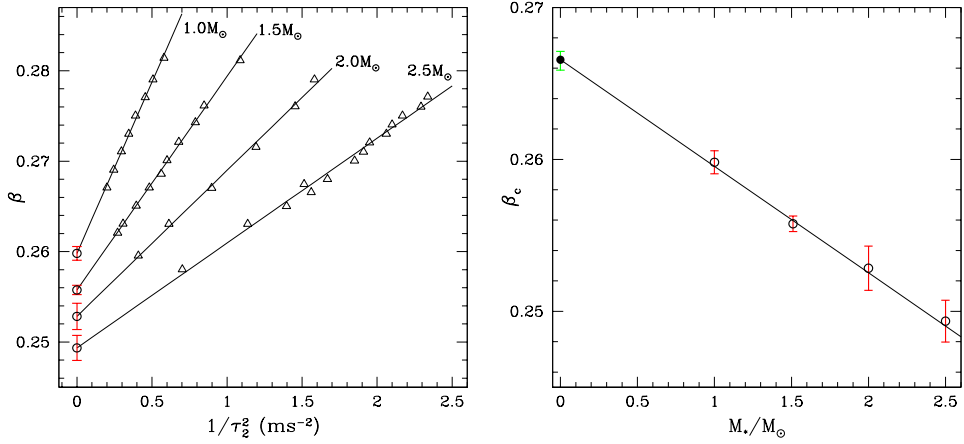

\begin{center}
\includegraphics[width=0.49\textwidth]{\imgname{critical_fit}}
%\hspace{-1cm}
\includegraphics[width=0.49\textwidth]{\imgname{estrapolazione_newtoniana}}
\end{center}
\vspace{-.7cm}
\caption{{\it Left panel:} Indicated with triangles are the data already
  reported in Fig.~\ref{fig:critical_fit} but shown here as a function of
  $1/\tau^2$ for the four constant--rest-mass sequences considered. The
  open circles indicate instead the extrapolation to $\beta_c$. The
  numerical values of the four fits are reported in
  \tref{table:fit_5_35_beta}. {\it Right panel:} The open circles are the
  same as in the left panel but shown as a function of the baryonic mass
  $M_*$; the filled circle, on the other hand, represents the limit of
  zero rest-mass and can be compared with the threshold value calculated
  in Newtonian gravity and with a linear stability analysis in
  ref.~\cite{KarinoEriguchi03}.}
\label{fig:NEW_EXTRAP}
\end{figure}

The left panel of Fig.~\ref{fig:NEW_EXTRAP} offers a different view of
the data already reported in Fig.~\ref{fig:critical_fit} by showing
$\beta$ as a function of $1/\tau^2$ for the four sequences considered
and thus highlighting the very good approximation in the
\textit{ansatz}~(\ref{eq:fit beta}). Indicated with open circles are
also the extrapolations of the critical value $\beta_c$ and these are
then reported in the right panel of Fig.~\ref{fig:NEW_EXTRAP} as a
function of the baryonic mass $M_*$ of the four sequences. Clearly,
they show a linear dependence on $M_*$, which can be expressed
phenomenologically as
\begin{equation}
  \label{eq:NEW_EXTRAP}
  \beta_c(M_*)=0.266(1)-0.0070(3) \left(\frac{M_*}{M_{\odot}}\right)\;,
\end{equation} 
and allows us to extrapolate the value of the threshold for in the
limit of a zero baryonic mass. Interestingly, the resulting number
$\beta_c(M_*=0)=0.266$ is in very good agreement with the value of
$\beta_c=0.266$ obtained in Newtonian gravity~\cite{KarinoEriguchi03}
but through a linear stability analysis for a sequence of equilibrium
models with the same polytropic index and degree of differential
rotation used here. This agreement represents an additional
confirmation of the accuracy and robustness of our extrapolation
method in determining the position of the threshold.

%%%----------------------------------------------------------------
%%% Fig -- Estrapolazione Newtoniana
%%%----------------------------------------------------------------
%%%----------------------------------------------------------------
%%% END Fig -- Estrapolazione Newtoniana
%%%----------------------------------------------------------------

Using the phenomenological dependence of the threshold $\beta_c$ on the
stellar rest mass given by expression~(\ref{eq:NEW_EXTRAP}), we have also
reconstructed the dependence of $\beta_c$ on the stellar compactness
$M/R_e$. In practice, for a large number of values of $M_*$ between $0$
and $2.5 M_\odot$ we have computed the value of the proper equatorial
radius $R_{e}$ and of the gravitational mass $M$ of the corresponding
stellar model in equilibrium and then used eq.~(\ref{eq:NEW_EXTRAP}) to
estimate the value of $\beta_c$. The result of this is shown as a solid
line in Fig.~\ref{fig:beta vs compactness}, with the dashed band
representing the estimated error obtained using the least-square fitting.
Furthermore, a good polynomial reconstruction of the median of the error
bar in Fig.~\ref{fig:beta vs compactness} suggests a quadratic dependence
of the threshold on the compactness, with coefficients given by
%%%
%%\begin{equation}
%%\label{betac_vs_MoR}
%%  \beta_c=0.266 - 0.18 \left(\frac{M}{R_e}\right)+
%%	  0.41\left(\frac{M}{R_e}\right)^2\;.
%%\end{equation}
%%%
%
\begin{equation}
\label{betac_vs_MoR}
  \beta_c=0.266 - 0.18 \left(\frac{M}{R_e}\right)+
	0.36\left(\frac{M}{R_e}\right)^2\;.
\end{equation}
Once perturbative calculations will be developed in the regime of
rapid and differential rotation considered here,
expression~(\ref{betac_vs_MoR}) can be used as a guideline to the
perturbative approach and the numerical measurements of the threshold
used to assess the validity and accuracy of the perturbative
approximation.

%%%----------------------------------------------------------------
%%% Fig -- Estrapolazione Newtoniana
%%%----------------------------------------------------------------
\begin{figure}[t]
\begin{center}
%\includegraphics[width=0.49\textwidth]{\imgname{compactstar}}
%\hspace{-1cm}
%\includegraphics[width=0.49\textwidth]{\imgname{saturation}}
\hspace{0.29\textwidth}
\includegraphics[width=0.69\textwidth]{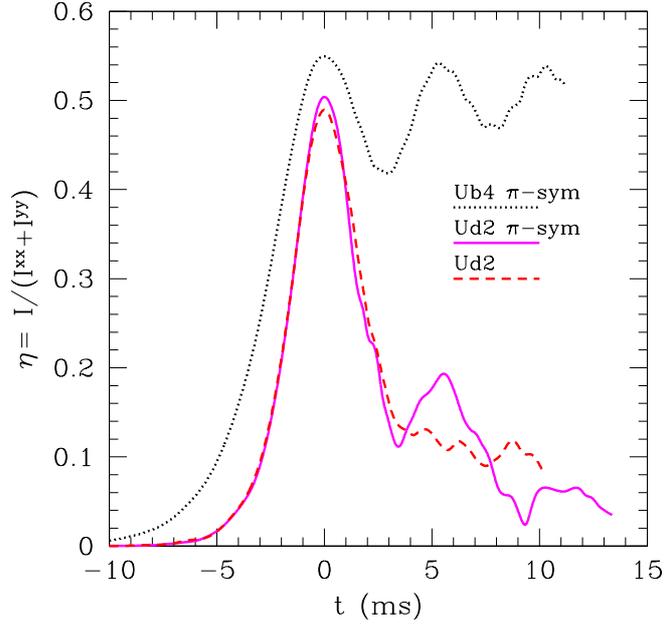}
\end{center}
\vspace{-.7cm}
\caption{Comparison of the evolution of the distortion parameters
  $\eta$ for a low-mass, low-compactness model Ub4 and a high-mass,
  high-compactness model Ud2. Indicated with a dotted and solid line
  are the simulations for Ub4 and Ud2 using a $\pi$-symmetry, while
  indicated with a dashed lines is the evolution for model Ud2 without
  $\pi$-symmetry. All the evolutions have been suitably shifted in
  time so as to have the maximum deformation at $t=0$.}
\label{fig:compactstar}
\end{figure}
%%%----------------------------------------------------------------
%%% END Fig -- Estrapolazione Newtoniana
%%%----------------------------------------------------------------

Overall, these results represent the first quantitative determination
of the dependence of the critical $\beta$ on the compactness of the
star for a selected profile of differential rotation and EOS. A
similar investigation was carried out also in ref.~\cite{Saijo:2001}
using the post-Newtonian (PN) approximation and rather large initial
$m\!\!=\!\!2$ perturbations with $\delta_2$=$0.1$. While the results
of that analysis had the limitation of the PN approximation and were
affected by the large initial perturbations, they also provided the
first evidence that the threshold is smaller for stellar models with
larger compactness.
 
%-----------------------------------------------------------------
\subsection{Persistence of the bar}
\label{sec:persistence}
%-----------------------------------------------------------------

As mentioned in the Introduction and for all of the models considered
here, any non-axisymmetric deformation that develops as a result of a
dynamical instability is also suppressed over a time comparable
(\textit{i.e.}, of the order of a few) dynamical timescales. In
addition, we have found that the persistence of the bar deformation
increases as $\beta - \beta_c$ tends to zero. Besides confirming what
already discussed in paper I, this behaviour matches the expectation
that the persistence of the bar is related to the degree of
overcriticality, with the duration of the saturation increasing as the
threshold is approached. For a star with a small compactness this time
would tend to the radiation-reaction timescale for a model with
$\beta=\beta_c$ and to zero for a model with $\beta \gg \beta_c$. For
a star with large compactness, on the other hand, the persistence near
the threshold can be further reduced by the stronger gravitational
fields.

Here, we also intend to gain insight on the role played by the stellar
compactness on the persistence of the bar deformations and, to this
scope, we have extended the time over which the simulations are
carried out for some selected models. Figure~\ref{fig:compactstar}
summarizes the results of these extended simulations by reporting the
evolution of the distortion parameter $\eta$ for two models,
\textit{i.e.}, Ub4 (dotted line) and Ud2 (solid line) having different
masses and compactnesses (\textit{i.e.}, $M=1.5\,M_{\odot}$,
$M=2.5\,M_{\odot}$ and $M/R_e\simeq 0.066$, $M/R_e\simeq 0.125$,
respectively), but with a similar distance from the threshold
(\textit{i.e.}, $\beta - \beta_c \simeq 0.007-0.008$). The two
evolutions have been suitably shifted in time so as to have the
maximum deformation at $t=0$.

Both of these models have been evolved using a $\pi$-symmetry in order to
remove the effect of the odd-mode coupling on the suppression of the bar
deformation and yet they show a remarkable difference.  The low-mass,
low-compactness model Ub4 reaches and maintains a rather large bar
deformation over several milliseconds of evolution. The high-mass, high-compactness
model Ud2, on the other hand, reaches comparable deformations but these
are rapidly suppressed over a few milliseconds, despite the use of the
$\pi$-symmetry. This behaviour underlines something that was already
remarked in paper I, namely that, depending on the specific stellar
properties, the bar deformation can be suppressed also by factors other
than the mode coupling which is most effective for models near the
threshold. In particular, for low-mass models (such as model U13 in paper
I) this effect is simply the excess of rotational kinetic which is
efficiently converted into internal one (see discussion in Sect. VIB of
paper I). For high-mass and high-compactness models (such as model Ud2
above), on the other hand, the strength of the gravitational field,
together with the excess kinetic energy, are very efficient in
suppressing the bar-deformation. This was not evident in the stellar
models considered in paper I, which all had $M_*=1.5\,M_{\odot}$.

To confirm that the $m\!\!=\!\!1$-mode coupling plays no significant
role in the dynamics of model Ud2, we show with a dashed line in
Figure~\ref{fig:compactstar} the evolution of the bar deformation for
model Ud2 in a simulation in which the $\pi$-symmetry was not
enforced. As it is obvious in the comparison between the solid and
dashed lines, the lack of symmetry does not change the suppression of
the bar-mode deformation and a difference emerges only at $t\sim 5$
ms, when the simulation with $\pi$-symmetry tends to revive the bar
deformation, while this does not happen in the simulation without the
$\pi$-symmetry.

%-----------------------------------------------------------------
\subsection{Unstable deformations with $m\!\!=\!\!3$}
\label{sec:unexpected}

%-----------------------------------------------------------------

While studying the dynamics of models with $M_*\ge 2M_{\odot}$ and
values of $\beta$ near the threshold for the development of the
bar-mode instability, we have also found stellar models that show the
development of a dynamical instability with a dominant $m\!\!=\!\!3$
deformation. Interestingly, this instability developed without the
introducing of any $m\!\!=\!\!3$ initial perturbation as was instead
done in ref.~\cite{Rampp:1998}. More precisely, we found four models
(Sc2, Sd6, Sd5, Sd4) for which the $m\!\!=\!\!3$ mode is the fastest
growing dynamically unstable deformation. In addition, we also see
evidence of a dynamical unstable growth of the $m\!\!=\!\!3$ mode for
models Sb1, Ub1, Sd3, Sd2, Sd1 and Ud1. However, because all the above
models lay very close to the threshold for the onset of the bar-mode
instability, nonlinear mode-couplings may be very important in this
region and the cause of the observed growths.

\begin{figure}[t]
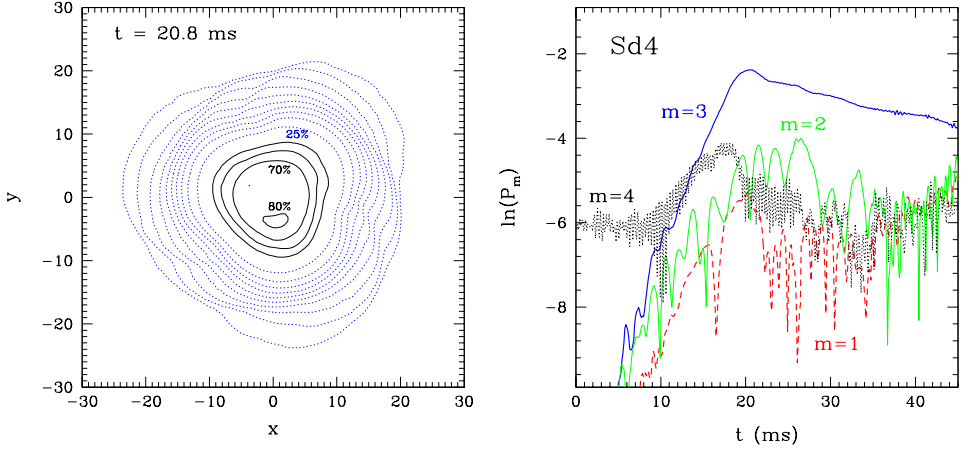

\begin{center}
\includegraphics[width=0.49\textwidth]{\imgname{245D_rho}}
%\hspace{-1cm}
\includegraphics[width=0.49\textwidth]{\imgname{245D_modesCM}}
\end{center}
\vspace{-.7cm}
\caption{{\it Left panel:} Rest-mass isodensity model Sd4. Starting from
  the center the solid lines represent respectively: 80\%, 70\%, 60\% and
  50\% of the maximum density. Dotted lines are $1/2^n$ times the maximum
  density, with $n=2,\ldots,11$. The snapshot was taken at t=20.8 ms.
  {\it Rigth panel:} Evolution of the global rest-mass density modes
  defined in eq.~(\ref{eq:modes}) for the same simulation. }
\label{fig:245D}
\end{figure}

The general behavior of the mode dynamics is reported in
Fig.~\ref{fig:245D} for the representative model Sd4, where in the
left panel we show the rest-mass isodensity contours in the equatorial
plane and at $t=20.8$ ms, while in the right one the evolution of the
global modes as defined in eq.~(\ref{eq:modes}) but as measured with
respect to the centre of mass of the system (see discussion in
Sect. IV of paper I for the definition of the centre of mass in this
context). When looking at this panel it is evident that the
$m\!\!=\!\!3$ unstable deformation is the fastest growing
one. Furthermore, the $m\!\!=\!\!2$ does not show any sign of an
unstable growth, which is instead shown by the $m\!\!=\!\!1$
deformation. Such a growth starts roughly at the time when the
$m\!\!=\!\!1$ and $m\!\!=\!\!2$ modes have powers comparable with the
background $m\!\!=\!\!4$, \textit{i.e.} at $t\sim 20$ ms, and it
continues for the rest of the simulation. The results on the
frequencies $f_3$ and the growth times $\tau_3$ of the $m\!\!=\!\!3$
instability are collected in Table~\ref{table:stable} and have been
computed using the same analysis discussed for the $m\!\!=\!\!2$
deformation.  It should be noted that the properties of the
$m\!\!=\!\!3$ instability are much more difficult to estimate
accurately as the corresponding deformations are smaller than those
measured for the bar-mode instability.

%%%----------------------------------------------------------------
%%% Fig -- MODE 3 UNSTABLE
%%%----------------------------------------------------------------
%%%----------------------------------------------------------------
%%% END Fig -- MODE 3 UNSTABLE
%%%----------------------------------------------------------------

It is probably worth stressing that this is the first time that
similar instabilities are seen in fully general relativistic
simulations of stellar models with a stiff EOS and a moderate degree
of differential rotation. In a recent work carried out in Newtonian
physics~\cite{2006ApJ...651.1068O}, in fact, the development of
deformations with $m\!\!=\!\!3$ was also found but for stellar models
with lower values of $\beta$ and with a stronger differential
rotation. Furthermore, because the models were studied by imposing an
initial artificial $m\!\!=\!\!2$ perturbation that could alter the
global dynamics, it is indeed difficult to understand if the origin of
this unstable deformation is purely dynamical or triggered by a
nonlinear coupling of modes. Nevertheless, it is interesting to note
that in all the simulations reported in~\cite{2006ApJ...651.1068O}
as well as in those presented here the $m$=$1$ deformation becomes the
dominant one in the final stages of the evolution. This lends support
to the idea that the suppression of non-axisymmetric deformations over
a dynamical timescale is a generic feature of these instabilities for
isolated stars and not necessarily restricted to stellar models with
high values of $\beta$.

At the moment and besides the classical $m$=$2$ bar-mode instability, a
proper understanding of the conditions that lead to the development of
dynamical instabilities is still lacking. The first attempts of
interpreting these instabilities, and especially the low-$\beta$
$m\!\!=\!\!1$ instability, have been made recently by several
authors~\cite{2006ApJ...651.1068O,Watts:2003nn}. However, further work is
needed to clarify the role of $\beta$, of the differential rotation law,
of the EOS and of the compactness in determining the growth times, the
maximum amplitudes and the persistence of these unstable deformations.
Most importantly, work will be needed to finally reach sufficient
conditions for the onset of these instabilities whose development has
been revealed by numerical simulations.

%XXXXXXXXXXXXXXXXXXXXXXXXXXXXXXXXXXXXXXXXXXXXXXXXXXXXXXXXXX%
%-----------------------------------------------------------------
\section{Conclusions}
\label{sec:conclusions}
%-----------------------------------------------------------------
%XXXXXXXXXXXXXXXXXXXXXXXXXXXXXXXXXXXXXXXXXXXXXXXXXXXXXXXXXXX%

We have presented accurate simulations of the bar-mode dynamical
instability in full General Relativity. The main motivation behind
this work was to address some important open questions about the
nonlinear features of non-axisymmetric dynamical instabilities in
rapidly rotating compact stars. The most important among these
questions, because of the impact it has on the global detectability of
these stars as sources of gravitational waves, is the determination of
the timescale over which the non-axisymmetric deformations persist
once these are fully developed.

In order to reach a better understanding of the physics governing
dynamical instabilities, we have analysed the onset and development of
the bar-mode instability for a large number of stellar models spanning
a wide range of masses and angular momenta. The initial models have
been calculated as stationary equilibrium solutions for axisymmetric
and rapidly rotating neutron stars modeled as relativistic polytropes
with adiabatic index $\Gamma=2$ and polytropic constant $K=100$. All the
stars have been constructed with a differential-rotation profile
having $\hat{A}=1$ and as members of four sequences of constant
rest-mass, with $M_*=1M_{\odot}$, $1.51M_{\odot}$, $2M_{\odot}$ and
$2.5M_{\odot}$, respectively. This large set of initial data
containing a total of $59$ models has allowed us to confirm and extend
the results presented in paper I. More precisely, we have analysed the
effects that the stellar compactness has on the threshold for the
onset of the dynamical bar-mode instability as well as of other
dynamical instabilities. Moreover, using an extrapolation technique
developed and tested in paper I, we have determined the threshold with
great accuracy and for a wide range of compactnesses, finding a good
agreement with the Newtonian prediction and improving the previous PN
estimates made in ref.~\cite{Saijo:2001}.

While studying the dynamics of models with $M_*\ge 2M_{\odot}$ and
values of $\beta$ near the threshold for the development of the
bar-mode instability, we have also found stellar models that show the
development of a dynamical instability with a dominant $m\!\!=\!\!3$
deformation, and without any seed perturbation of that type. The
appearance of these instabilities, whose growth-time and frequency
have been computed using the same methodology developed for the
bar-mode instability, may be rather generic in stars with high mass
and deserve additional attention.

Finally, we remark that for all the simulated models, the deformations
generated by the non-axisymmetric instabilities are suppressed over a
dynamical time-scale either as a result of nonlinear mode-couplings or
as a result of the conversion of the excess rotational kinetic energy
into internal one. In all cases we have observed the emergence of a
residual $m$=$1$ deformation in the final stages of the instability
and before an axisymmetric configuration is recovered. These results
confirm our previous findings presented in paper I and are in
agreement with those recently reported in~\cite{2006ApJ...651.1068O},
thus lending support to the idea that the suppression of
non-axisymmetric deformations over a dynamical timescale is a generic
feature in isolated stars. Further work is clearly needed to confirm
or reject this conjecture and to derive the sufficient conditions for
the onset of these ``odd-$m$'' instabilities.

\ack 

It is a pleasure to thank Harald Dimmelmeier and Ian Hawke for useful
discussions and suggestions. GMM is also grateful to Jan Christian
Bryne and the Bergen Center for Computational Science for the kind
hospitality. Support for this research comes also through the SFB-TR7
of the German DFG and through the OG51 of the Italian INFN. GMM is
presently a ``Dalla Riccia'' fellow at the MPA.  All the computations
were performed on the cluster for numerical relativity
\textit{``Albert''} at the University of Parma and \textit{``Peyote''}
at the AEI.

%\bibliographystyle{unsrt}
%\bibliography{Compactness2006}

\section*{References}

\begin{thebibliography}{10}

\bibitem{Baiotti:2006wn}
   Baiotti L, De~Pietri R, Manca G M and Rezzolla L 2007 
   \PR {\bf D 75}, 044023.
%
%\newblock Accurate simulations of the dynamical barmode instability in full
%  {G}eneral {R}elativity.
%\newblock {\em Submitted to Phys. Rev. D}, 2006.

\bibitem{ShibataSekiguchi2004}
   Shibata M and  Sekiguchi Y 2005 \PR {\bf D 71}, 024014.
%\newblock Three-dimensional simulations of stellar core collapse in full
%  general relativity: {N}onaxisymmetric dynamical instabilities.
%\newblock {\em Phys {R}ev.}, D71:024014, 2005.
%\bibitem{Saijo2005}
\item[]
   Saijo M 2005, \PR {\bf D71}, 104038.
%M.~{Saijo}.
%\newblock Dynamical bar instability in a relativistic rotational core collapse.
%\newblock {\em Phys. {R}ev.}, D71(10):104038, 2005.
%\bibitem{Ott05}
\item[]
   Ott C D, Ou S, Tohline J E and Burrows A 2005
   {\it Astrophys. {J}.} {\bf 625}, L119--L122.
%Christian~D. Ott, Shangli Ou, Joel~E. Tohline, and Adam Burrows.
%\newblock One-armed {S}piral {I}nstability in a {L}ow {T}/|{W}| {P}ostbounce
%  {S}upernova {C}ore.
%\newblock {\em Astrophys. {J}.}, 625:L119--L122, 2005.
%\bibitem{Ott:2006eu}
\item[]
  Ott C. D. \etal 2006 {\it ``3D Collapse of Rotating Stellar Iron Cores in 
 General Relativity with Microphysics''}, arXiv:astro-ph/0609819v1.

%\newblock 3{D} {C}ollapse of {R}otating {S}tellar {I}ron {C}ores in {G}eneral
%  {R}elativity with {M}icrophysics.
%\newblock 2006.

%%----------------------------

\bibitem{2006ApJ...644.1063D}
  {Dessart} L, {Burrows} A, {Ott} C D, {Livne} E, {Yoon} S C and
  {Langer} N 2006
  %\newblock {{M}ultidimensional {S}imulations of the {A}ccretion-induced
  %{C}ollapse of {W}hite {D}warfs to {N}eutron {S}tars}.
  {\it \apj} {\bf 644}, 1063.
  %%--1084, June 2006.

%\bibitem{Shibata:2006nm}
%%%%
\bibitem{Shibata:2005ss}
  Shibata M, Taniguchi K and Uryu K 2003
  %\newblock Merger of binary neutron stars of unequal mass in full general
  %relativity.
  \PR {\bf D 68}, 084020.
\item[]
  Shibata M, Taniguchi K and Uryu K 2005
  %\newblock Merger of binary neutron stars with realistic equations of state in
  %full general relativity.
  \PR {\bf D 71}, 084021.
\item[]
  Shibata M and Taniguchi K 2006
  %\newblock Merger of binary neutron stars to a black hole: {D}isk mass, short
  %gamma-ray bursts, and quasinormal mode ringing.
  \PR {\bf D 73}, 064027.
%%
%%\bibitem{Shibata:2003ga}


\bibitem{Bonaldi:Dual2003}
        Bonaldi M \etal 2003 \PR {\bf D 68}, 102004.
\item[] Bonaldi M \etal 2006 \PR {\bf D 74}, 022003.


\bibitem{2006ApJ...651.1068O}
  {Ou} S and  {Tohline} J E  2006
  %\newblock {{U}nexpected {D}ynamical {I}nstabilities in {D}ifferentially
  %{R}otating {N}eutron {S}tars}.
  {\it \apj} {\bf 651}, 1068.


\bibitem{Stergioulas95}
   Stergioulas N and Friedman J L 1995
   %\newblock Comparing {M}odels of {R}apidly {R}otating {R}elativistic {S}tars
   %{C}onstructed by {T}wo {N}umerical {M}ethods.
   {\it Astrophys. {J}.} {\bf 444}, 306.

\bibitem{Richtmyer67}
 Richtmyer R D and Morton K W 1967
\newblock {\it Difference {M}ethods for {I}nitial {V}alue {P}roblems}.
\newblock Interscience Publishers, New York, 1967.

\bibitem{Nakamura:87}
   Nakamura T, Oohara K and Kojima Y 1987
   {\it Prog. {T}heor. {P}hys. {S}uppl.}, {\bf 90}, 1.

\bibitem{Baiotti03a}
  Baiotti L, Hawke I, Montero P and Rezzolla L 2003
  \newblock A new three-dimensional general-relativistic hydrodynamics code.
  in R.~Capuzzo-Dolcetta, editor, {\it Computational {A}strophysics in
  {I}taly: {M}ethods and {T}ools}, volume~1, page 327, Trieste, 2003. Mem. Soc.
  Astron. It. Suppl.
%%\bibitem{Baiotti04a}
\item[]
  Baiotti L \etal 2005
  %Baiotti L, Hawke I, Montero P J, L{\"o}ffler F, Rezzolla L,
  %Stergioulas N, Font J A and Seidel E. 2005
  %Luca Baiotti, Ian Hawke, Pedro~J. Montero, Frank L{\"o}ffler, Luciano Rezzolla,
  %  Nikolaos Stergioulas, Jos\'e~A. Font, and Ed~Seidel.
  %\newblock Three-dimensional relativistic simulations of rotating neutron star
  %  collapse to a {K}err black hole.
  %\newblock {\em Phys. {R}ev.}, 
  \PR {\bf D 71}, 024035..

\bibitem{Chandrasekhar69}
 Chandrasekhar S 1969
 \newblock {\it Ellipsoidal {F}igures of {E}quilibrium}.
 \newblock Yale Univ. Press, New Haven.

\bibitem{lomb76}
  Lomb N R  1976
  %\newblock Least-squares frequency analysis of unequally spaced data.
  %\newblock 
  {\it Astr. {S}pace {S}ci.}, 39:447--462, 1976.

\bibitem{KarinoEriguchi03}
  Karino Sand Eriguchi Y 2003
  %\newblock Linear {S}tability {A}nalysis of {D}ifferentially {R}otating
  %{P}olytropes: {N}ew {R}esults for the m=2 f-{M}ode {D}ynamical {I}nstability.
  %\newblock 
  {\it \apj} {\bf 592}, 1119--1123., 2003.

\bibitem{Saijo:2001}
  Saijo M, Shibata M, Baumgarte T W and Shapiro S L 2001
  %Motoyuki Saijo, Masaru Shibata, Thomas~W. Baumgarte, and Stuart~L. Shapiro.
  %\newblock Dynamical {B}ar {I}nstability in {R}otating {S}tars: {E}ffect of
  %{G}eneral {R}elativity.%\newblock 
  {\it Astrophys. {J}.} {\bf 548}, 919--931.

\bibitem{Rampp:1998}
   Rampp M, Mueller E and Ruffert M 1998
  {\it Astron. {A}strophysics}, {\bf 332}, 969.

\bibitem{Watts:2003nn}
   Watts A L, Andersson N, and Jones D I 2005
   {\it Astrophys. {J}.} {\bf 618}, L37.
%\bibitem{saijo2006MNRAS}
\item[]
  {Saijo} M and Yoshida S 2006
  %\newblock {{L}ow {T}/|{W}| dynamical instability in differentially rotating
  % stars: diagnosis with canonical angular momentum} \newblock 
  {\it Mon. {N}ot. {R}oy. {A}stron. {S}oc.} {\bf 368}, 1429--1442.

\end{thebibliography}

\end{document}